\documentclass[aps,prb,preprint,superscriptaddress,showpacs,amsmath,amssymb,floatfix,tighten,nofootinbib]{revtex4-2}
\usepackage{graphicx}
\usepackage{bm}
\usepackage{threeparttable}
\usepackage{multirow}
\usepackage[table]{xcolor}
\usepackage{lipsum,multirow}
\usepackage{amssymb,amsmath,epsfig,rawfonts}

\usepackage{float}
\usepackage{color}
\definecolor{dgreen}{RGB}{0,204,0}
\usepackage{mathptmx}
\usepackage[breaklinks=true,colorlinks=true,urlcolor=blue, citecolor=blue,linkcolor=blue,bookmarks=false]{hyperref}

\graphicspath{{figs/}}

\begin{document}

\title{Surface termination dependence of electronic and optical properties in Ti$_2$CO$_2$ MXene monolayers}

\author{Zafer Kandemir}
\affiliation{Department of Mechanical Engineering, Faculty of Engineering, Eskisehir Technical University, 26555, Eskisehir, Turkey}
\author{Engin Torun}
\thanks{Current Adress: Simbeyond B.V., Het Eeuwsel 57, AS Eindhoven 5612, Netherlands}
\noaffiliation{}
\author{Fulvio Paleari}
\affiliation{Istituto di Struttura della Materia and Division of Ultrafast Processes in Materials (FLASHit) of the National Research Council, via Salaria Km 29.3, I-00016 Monterotondo Stazione, Italy.}
\author{Celal Yelgel}
\affiliation{Department of Electricity and Energy,  Recep Tayyip Erdogan University, 53100, Rize, Turkey}
\author{Cem Sevik}
\affiliation{Department of Mechanical Engineering, Faculty of Engineering, Eskisehir Technical University, 26555, Eskisehir, Turkey}
\email{csevik@eskisehir.edu.tr}
\date{\today}

\begin{abstract}
Two-dimensional (2D) MXenes are a rapid growing family of 2D materials with rich physical and chemical properties where their surface termination plays an essential role. Among the various 2D MXenes, functionalization of the Ti$_{n}$C$_{n-1}$ phase with oxygen (O) atoms makes them attractive for optoelectronic applications due to their optical gap residing in the infrared or visible region. In this manuscript, we theoretically investigate the electronic and optical properties of four different O-atom-functionalized Ti$_{n}$C$_{n-1}$  MXene monolayers using state-of-the-art, first-principles techniques. In particular, we calculate the quasiparticle corrections on top of density functional theory (DFT) at the GW level and the exciton-dominated optical spectra by solving the Bethe-Salpeter equation (BSE) also at finite momentum. We find that all but one of the monolayer models are indirect band gap semiconductors where quasiparticle corrections are very important ($\sim 1$ eV). The optical spectra are instead dominated by direct and indirect excitons with large binding energies (between $0.5$ and $1$ eV).
Most direct excitons lie above $1.5$ eV, while the indirect ones are below: therefore, we conclude that Ti$_{n}$C$_{n-1}$ should display strong absorption in the visible region, but phonon-assisted emission in the infrared. Our work thus reveals the potential usage of surface terminations to tune the optical and electronic properties of Ti$_{n}$C$_{n-1}$ MXene monolayers, while emphasizing the pivotal role of many-body effects beyond DFT to obtain accurate prediction for these systems.
\end{abstract}


\maketitle

\section{Introduction}
The family of 2D transition metal carbide, nitride, and carbonitride materials -- the so-called ``MXenes" -- possessing chemical formula of M$_{n+1}$X$_n$T$_y$ ($n$=1, 2 or 3), where ``M" is an early transition metal such as Sc, Ti, Zr, Hf, V, Nb, Ta, Cr, Mo or W, ``X" is either N or C, and ``T$_y$" stands for the surface terminations such as O, OH, F or S has been the object of great interest since its first introduction by Gogotsi et al.\cite{naguib2011} These layered materials are mostly chemically synthesized through selective acid etching of A elements from MAX phases, \cite{doi:10.1021/acsnano.9b06394,Champagne_2020,naguib2014,Venkateshalu2020} where "A" is a group IIIA to VIA element. In addition, chemical transformations and bottom-up construction techniques\cite{anasori20172d} such as chemical vapor deposition have been also demonstrated for the successful synthesis of some MXene crystals. To date, many kinds of MXene crystals have been experimentally realized\cite{anasori20172d,nguyen2020,wang2020mxenes,lei2015recent,wang2020} and methods to control formation on surface termination have been demonstrated as well. \cite{Kamysbayev2020,hart2019control} Indeed, large-scale single-layer Ti$_2$CO$_2$ crystals have not been reported along with proper characterization of a physical property such as optical. However, due to the enormous research effort and current developments on techniques for MXene Delamination into Single-Layer Flakes it is close to coming true\cite{hawking1988}. On the other hand, single-layer crystals with random functional groups such as OH, F, O, Cl are available in the literature\cite{6b03064}.

Numerous different MXene layered systems arise by combining their chemical versatility and thier wide range of surface functionalizations, as demonstrated in research studies revealing the enormous potential of MXenes in various applications\cite{wang2020mxenes,lei2015recent,VahidMohammadi2021} such as power generation and storage, \cite{anasori20172d,wang2020,PhysRevApplied.12.014001,sevik1,C4CP00467A} gas, piezoresistive and strain sensors, \cite{LeeLee1,YananYanan1,CaiCai1,Zhangeaat0098,PhysRevMaterials.2.074002} chemical catalysis, \cite{nguyen2020} water purification, \cite{zhang2016compu,Srimuk,RenRen} plasmonics, \cite{JaksicJaksic,SarychevaSarycheva,ChaudhuriChaudhuri} transparent conductors\cite{hantanasirisakul2016fab} and electromagnetic interference shielding.\cite{Shahzad1137} Recent studies have also suggested the occurrence of superconductivity and of magnetic properties which might be the subject of future qubit and skyrmion-based investigations. \cite{Kumar1,Kamysbayev2020,Xu2015NatMa,D0NR03875J}

Among the experimentally available MXenes, Ti$_n$C$_{n-1}$T$_n$ is the most widely investigated one due to its largest superficial area per weight and being one of the thinnest MXene phases. \cite{articleMxene} First-principles calculations have shown that the pristine Ti$_{n}$C$_{n-1}$ are metallic. \cite{Champagne_2020} However, after functionalization with O, 2D Ti$_2$C becomes semiconducting with a considerable band gap energy, \cite{zhang2019prediction} this  also holds for the Zr$_{2}$CO$_2$ and Hf$_2$CO$_2$ phases, as well.\cite{C7CP02513K}

The notable influence of the surface termination on the physical properties, e.g., electronic, mechanical, ionic diffusion, and ionic absorption have already been investigated for O-terminated Ti$_2$C monolayers.\cite{Zha_2015,Khazaei2013novele,Khazaei20142dmo2c,Xie2014role} The effect of surface termination on the optical properties of these materials, on the other hand, has not been systematically investigated and the literature is rather sparse. For instance, the optical gap and the binding energy of the corresponding first bright direct exciton have been reported using GW and BSE formalism for the most chemically stable O-terminated Ti$_2$C monolayer as 1.26 eV and 0.56 eV, respectively. \cite{ding2020many} The absorbed photon flux has been calculated as 1.76 mAcm$^{-2}$  which is comparable with the 1 nm thick layers of Si (0.1 mAcm$^{-2}$), GaAs (0.3  mAcm$^{-2}$), and P3HT polymer (0.2  mAcm$^{-2}$). \cite{ding2020many, doi:10.1021/nl401544y} This consequently emphasizes the potential of using these monolayers in photodetection and photovoltaic applications. On the other hand, most of the semiconductor MXene structures have been determined as indirect band gap semiconductors, \cite{zhang2019prediction,C7CP02513K} which means that it is important to include indirect excitons in the computational studies to predict the optical properties and application areas of the different MXene structures. Taking into consideration all these facts, in this manuscript we present a thorough analysis based on first-principles calculations on the ground and excited state properties of the four different O terminated models of Ti$_2$CO$_2$ monolayers. We first present the electronic structures of the monolayer models and show that all of them but one are indirect band gap semiconductors at both DFT and GW levels. Subsequently, we investigate the optical properties of these monolayers including direct and indirect excitons by solving BSE using quasi particle (QP) energies and DFT wave functions. We observe that the bound indirect excitons are the lowest lying ones in the absorption spectra of indirect monolayer models although their binding energies are in average lower than the direct ones.

This manuscript is organized as follows: we first give the computational details in Sec. \ref{Comp_met}.  Then in Sec. \ref{Results} we present and discuss the results corresponding to the DFT, GW, and BSE calculations. Finally, we summarize our main findings in Sec. \ref{SunCon}.

\section{Computational methods}\label{Comp_met}

The DFT ground state calculations were performed with Quantum ESPRESSO\cite{qe1,qe2} using the local density approximation\cite{reflda} (LDA) and norm-conserving pseudopotentials. \cite{dojo} The energy cutoff for the plane wave basis was set to $90$ Ry and a $\Gamma$-centered 18$\times$18$\times$1 k-point mesh was used, which guarantees a total energy convergence of 10$^{-10}$ eV during the self-consistent steps. The geometries were fully optimized until the Hellmann-Feynman forces on each atom were less than $0.02$ eV/{\AA}.
A vacuum separation of 20 {\AA} in the out-of-plane direction was ensured to eliminate spurious interactions between monolayers with their periodically repeated images. Spin-orbit coupling is not taken into account in the simulations presented in this manuscript as our tests revealed its effects being negligible for the systems under investigation.

The many-body perturbation theory (MBPT) calculations, performed on top of the DFT results, were conducted with the YAMBO code.\cite{yambo1,yambo2}
The G$_0$W$_0$\cite{ref_gw} corrections to the Kohn-Sham eigenvalues were computed with a plasmon-pole approximation for the dynamical electronic screening.
The direct and indirect band gaps were converged with a 48 $\times$ 48 $\times$ 1 $k$-grid mesh, 217 k points in the irreducible Brillouin zone (BZ), summing over $400$ states for both the screening function and the Green's function.  The corrections were computed for the top 3 valence bands and the bottom 3 conduction bands.
The BSE\cite{bse-ref2} was then solved with RPA static screening, which was summed over $400$ bands and in the Tamm-Dancoff approximation on top of the GW results.
The direct and indirect (i.e. finite-momentum) exciton energies and their wave functions were obtained for the first 5500  excitonic states by using the iterative scheme enabled by the SLEPC library.\cite{slepc} The Coulomb cutoff (CC) technique was used along the out-of-plane direction to eliminate the interactions to all the undesired images of the systems in both G$_0$W$_0$ and BSE steps.\cite{Ismail2006} Convergence tests for the parameters used in MBPT calculations are provided in the supplementary material\cite{smaterials}. \footnote{The input parameters were individually converged by slowly increasing them until differences in band gaps and exciton energies between two subsequent runs were below 0.1 eV}

\begin{figure}[h!]
\centering
{\includegraphics[width=0.9\linewidth]{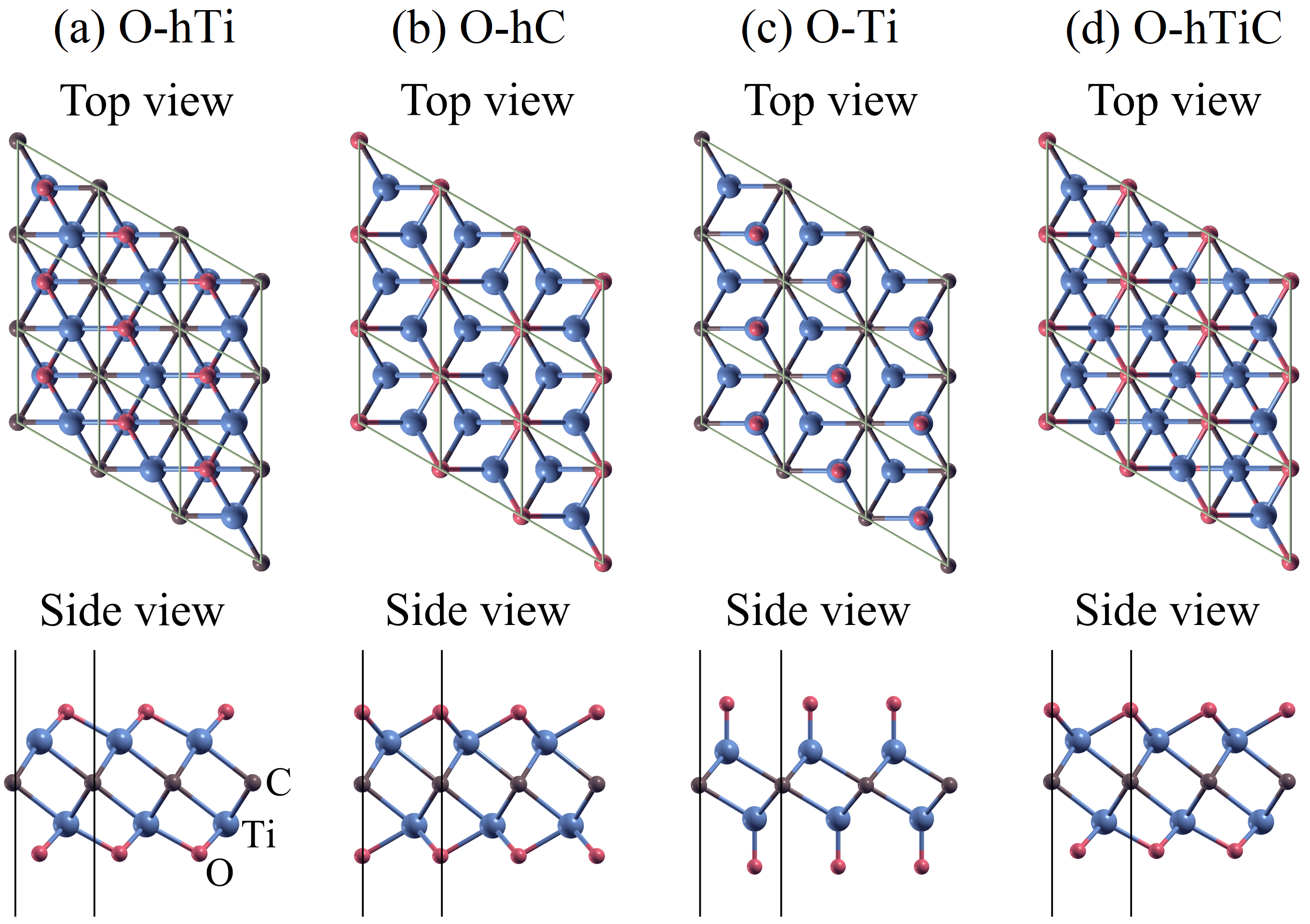}}
\caption{(Color online)
Top and side views of the optimized crystal structures of the O-terminated Ti$_2$CO$_2$ monolayers. A $3\times3$ supercell is shown for clarity. The blue, red and brown spheres represent titanium, oxygen and carbon atoms, respectively. (a) O-hTi; O atoms in the hollow site between Ti atoms and on top of the Ti atom in the opposite layer. (b) O-hC; O atoms again in the hollow site, but this time both of them are on top of the C atom. (c) O-Ti; both O atoms on Ti atoms. (d) O-hTiC; O atoms in the hollow sites, but with one O atom above the C atom and the other O atom above the opposite Ti atom.}
\label{fig1}
\end{figure}

\section{Results and Discussion}\label{Results}

Crystal structure of the four possible O-terminated hexagonal Ti$_2$C monolayer models are shown in Fig. \ref{fig1}(a-d). The corresponding binding energies of the O atom are predicted via the following equation:
\begin{equation}
E_b=\dfrac{1}{N}\big[E_{\mathrm{Ti}_{2}\mathrm{CO}_2}-E_{\mathrm{Ti}_{2}\mathrm{C}}-2E_{\mathrm{O}}]
\end{equation}
where $E_{\mathrm{Ti}_{2}\mathrm{CO}_2}$, $E_{\mathrm{Ti}_{2}\mathrm{C}}$, and $E_{\mathrm{O}}$ are the total energies of Ti$_2$CO$_2$, Ti$_2$C, and the isolated O atom, respectively ($N$ being the total number of atoms in the unit cell).

The calculated binding energies ($E_b$) along with lattice constants ($a$) and thicknesses ($d$) of all the investigated Ti$_2$CO$_2$ monolayers are reported in Table \ref{table1}. The results are in good agreement with recent works based on DFT calculations with different functionals.\cite{ding2020many,zhang2019prediction,zhang2016compu,bai2016dependence} Here, more negative binding energy indicates the more favorable exothermic binding of O atoms. Therefore, the most and the least chemically stable structures are predicted as O-hTi and O-Ti, respectively. We should note that these results are only to compare the chemical stability of these structures in their pristine form and any one of them could be stabilized by temperature, pressure, growth conditions and substrate effects.

\begin{table*}
\caption{Calculated parameters of the O-terminated Ti$_2$CO$_2$ monolayers: Lattice constant ($a$), monolayer thickness ($d$), binding energy of the O atom ($E_b$), band gap energies from the LDA ($E_{gap}^{LDA}$) and GW calculation ($E_{gap}^{GW}$), location of the valence band maximum (VBM), conduction band minimum (CBM) in the BZ and type of the band gap. Note that the minimum direct band gap of the indirect monolayers are reported in the paranthesis.}\label{table1}
\begin{threeparttable}
\begin{ruledtabular}
\begin{tabular}{lcccccccc}
System & $a$ ({\AA}) & $d$ ({\AA}) & $E_b$ (eV/atom)  & $E_{gap}^{LDA}$ (eV) & $E_{gap}^{GW}$ (eV) &   VBM      & CBM          & Type \\
\hline
\vspace{-0.3cm}
\\
O-hTi  &   2.98      &    4.35     &    -4.97         &       0.28(0.61)     &     1.29(1.86)      & $\Gamma$   &   $M$        & Indirect \\
\hline
\vspace{-0.28cm}
\\
O-hC   &   2.90     &     4.75     &    -4.59         &       0.36(0.61)     &     1.19(1.42)      & $K-\Gamma$ &  $\Gamma-M$  & Indirect\\
\hline
\vspace{-0.28cm}
\\
O-Ti   &  3.28      &     5.43     &    -3.83         &       0.68           &     1.92            &   $M$      &   $M$        & Direct\\
\hline
\vspace{-0.28cm}
\\
O-hTiC & 2.95       &     4.51     &    -4.79         &       0.74(1.30)     &     1.74(2.46)      & $\Gamma$   &   $M$        & Indirect\\
\end{tabular}
\end{ruledtabular}
\end{threeparttable}
\end{table*}

\begin{figure}
\centering
{\includegraphics[width=0.9\linewidth]{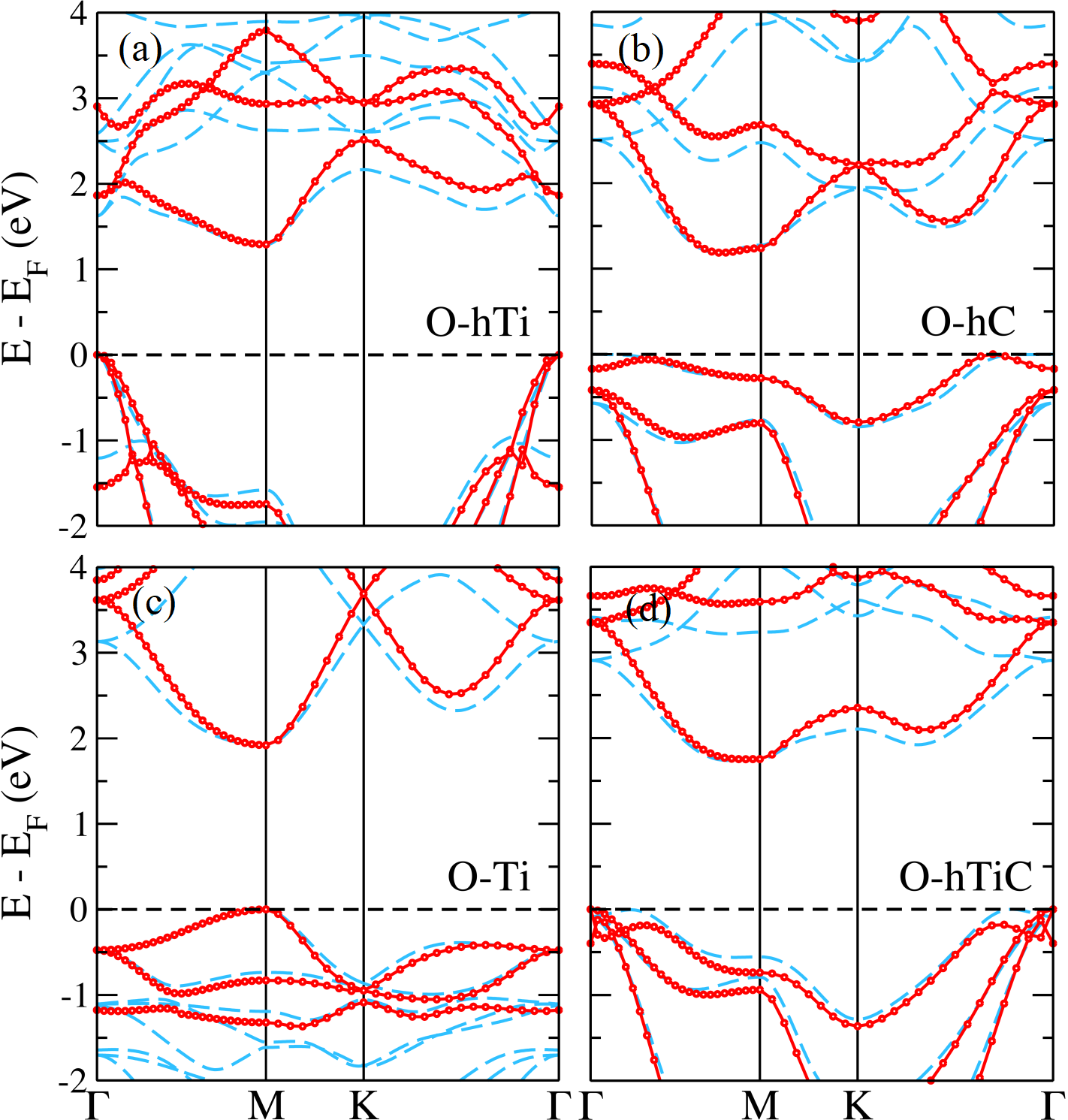}}
\caption{(Color online)
  Band structures of Ti$_2$CO$_2$ monolayers: (a) O-hTi, (b) O-hC, (c) O-Ti and (d) O-hTiC. The light blue dashed and red dotted lines represent LDA and G$_0$W$_0$ band structures, respectively. The LDA band structures are shifted to the GW band gap to compare the band dispersions. The black dashed line indicates the Fermi level which is shifted to 0 eV.}
\label{fig2}
\end{figure}

\subsection{Electronic structure and quasiparticles}
The pristine bulk Ti$_2$C (non-terminated) is metallic with a high density of states at the Fermi level.\cite{wang2014290} However, it turns into a semiconductor when terminated with O atoms. In order to address the DFT-LDA band gap underestimation which leads to discrepancies between calculated and experimental spectra, \cite{ding2020many,zhang2019prediction} we performed GW calculations to access the QP spectral properties. Fig.\ref{fig2} demonstrates the LDA and GW-corrected band structures of the Ti$_2$CO$_2$ monolayers. It is important to note that LDA gaps are shifted to the GW ones to compare the dispersion of the bands. It is observed that band dispersions are very similar at both LDA and GW levels for the valence bands but slightly different for the conduction bands particularly along $K-\Gamma$ direction. At both LDA and GW level, O-hTi, O-hC, and O-hTiC monolayers are found to be indirect but O-Ti a direct band gap semiconductor with a GW (LDA) band gap of 1.29 (0.28), 1.19 (0.36), 1.74 (0.74) eV, and 1.92 (0.68), respectively as reported
in Tab.~\ref{table1} together with the other properties of the monolayers. As can be seen that the band gap values are enormously increased by the self-energy corrections and
brings them much closer to the low-energy side of the visible spectrum.  In particular, the indirect QP band gap of O-hTi is now 1.29 eV, more than four times its DFT value, while the direct O-Ti gap is now 1.92 eV, increasing by almost three times. Our results are in good agreement with the available results for O-hTi structure, 1.32\cite{ding2020many} and 1.15\cite{zhang2019prediction} eV. The indirect band gap of O-hTi is 0.28 eV at the LDA level also agrees well with the reported ones which are in the range of 0.24 and 0.32 eV. \cite{zhang2016compu, xie2013, zhang2019prediction, ding2020many, hong2019, zha2015}

\begin{figure}
\centering
{\includegraphics[width=0.8\linewidth]{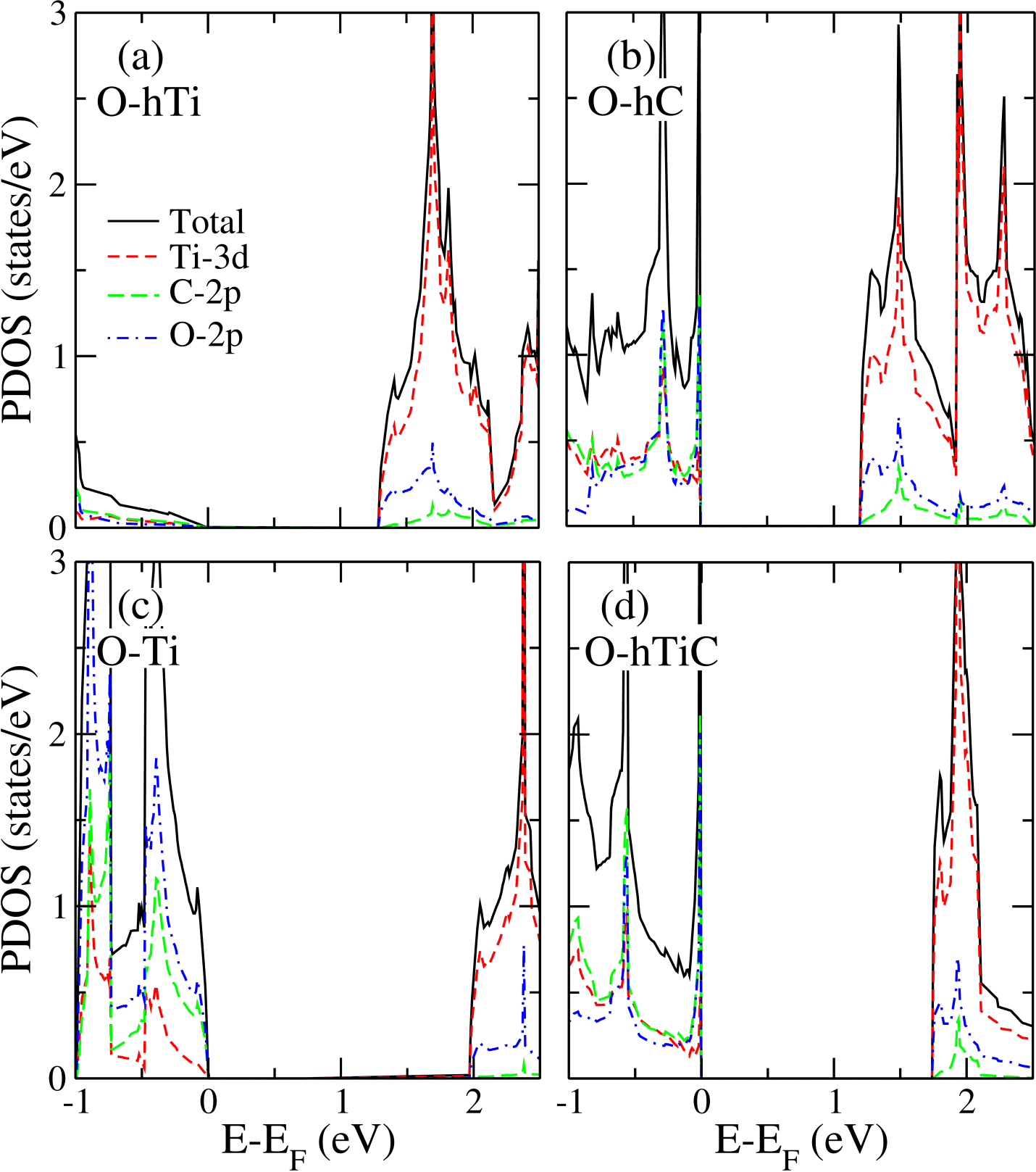}}
\caption{(Color online)
  Total and partial densities of states (DOS) of Ti$_2$CO$_2$ monolayers: (a) O-hTi, (b) O-hC, (c) O-Ti and (d) O-hTiC. The total DOS, and partial DOSes of Ti-3d, C-2p
  and O-2p orbitals are shown in black, red short-dashed, green long-dashed and blue dotted-dashed lines, respectively. Fermi energy is shifted to 0 ev.}
\label{fig3}
\end{figure}

Although having exactly the same atomic composition, the GW correction to the LDA band gap varies and resulting QP gaps do not follow the same energy ordering as in the DFT case
due to the different screening environment for the charge carrier interaction in each monolayer. The largest correction to the band gap, 1.24 eV, is calculated for the O-Ti monolayer, which might be expected as (i) the higher localisation of the isolated O orbitals leads to stronger corrections, and (ii) O-Ti is the thinnest one among the considered systems, therefore the electrons here are even more weakly screened compared to the other cases, again leading to larger band gap openings.

\begin{figure*}
\centering
{\includegraphics[width=0.9\linewidth]{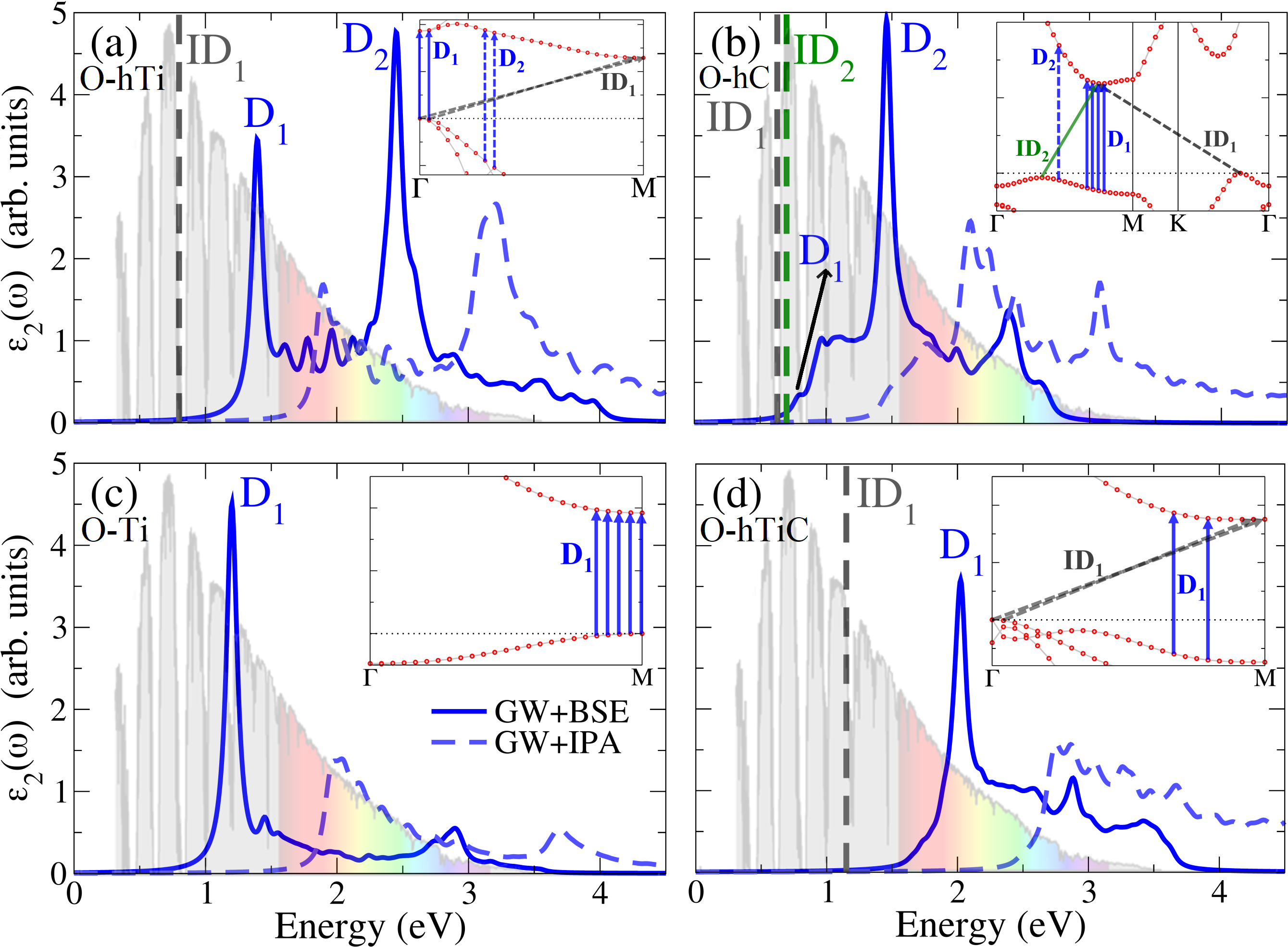}}
\caption{(Color online)
The imaginary part of the dielectric functions - proportional to the absorption spectrum - of the O-terminated Ti$_2$CO$_2$ monolayer models: (a) O-hTi, (b) O-hC, (c) O-Ti and (d) O-hTiC. The blue solid and blue dashed lines represent the spectrum computed with GW+BSE and GW+IPA methods, respectively. The solar flux of terrestrial (AM1.5g) spectra and visible-light region is shown in the background\cite{solar}. The energy positions of the lowest-lying finite-momentum excitons are shown with vertical dashed lines for comparison. The exciton states are labelled as in the main text. Insets: location in the BZ of the most relevant electron-hole contributions to the labelled excitons.}
\label{fig4}
\end{figure*}

Partial DOS analysis (Fig.\ref{fig3}) reveals that the 3d orbitals of Ti and 2p orbitals of C and O atoms partially contribute to the valence and
conduction bands of Ti$_2$CO$_2$ monolayers. Rather large DOS around Fermi level and conduction band minimum of the O-hC, O-Ti and O-hC monolayers can be noticed in the figure.
This ultimately leads to high joint DOS which is an indication of having strong light absorption and emission properties of these monolayers. On the other
hand, the very dispersive valence bands of O-hTi monolayer leads to drastically reduced DOS around Fermi level. This suppresses the joint DOS and hence the optical
response of the monolayer.

In view of these striking results, it clearly becomes necessary to investigate the effects of electron-hole interaction in order to accurately determine the absorption
and emission energy ranges of O-terminated layered Ti$_2$CO$_2$ systems.

\subsection{Optical properties and excitons}
The imaginary part, $\epsilon_2(\omega)$, of the frequency dependent dielectric function
$( \varepsilon(\omega) = \varepsilon_1(\omega) +i\varepsilon_2(\omega) )$
proportional to the optical absorption spectrum which is defined in the independent-particle approximation (IPA) as \cite{FulvioPhDThesis}
\begin{equation}
\varepsilon_2 (\omega) = \dfrac{8 \pi^{2}e^{2}}{V} \sum_{\kappa} | d_{\kappa} |^{2} \delta(\omega - \Delta \epsilon_{\kappa})
\label{eqn1}
\end{equation}
Here, $d_{\kappa}$ is the dipole matrix element and $\Delta \epsilon_{\kappa}$ is the transition energy of the electrons that absorb the incoming
electromagnetic field with frequency $\omega$. When the transition is allowed by the symmetry, a peak appears in the absorption spectrum
at the transition energy with an intensity proportional to the transition probability.

This approach might provide reasonable absorption spectra for the materials where the Coulomb interaction is highly screened and hence electron-hole interaction
has negligible effect on the optical response of the material. However, for the very thin materials, such as O-terminated Ti$_2$CO$_2$ monolayers, the screening
of the Coulomb interaction is drastically reduced due to the absence of screening in vacuum which in turn enhances the electron-hole interaction. Therefore,
it is necessary to plug in the electron-hole interaction into $\epsilon_2(\omega)$ (Eqn. \ref{eqn1}) via MBPT as
\begin{equation}
\varepsilon_2 (\omega) = \dfrac{8 \pi^{2}e^{2}}{V} \sum_{\lambda} | \sum_{\kappa} \bar{A}^{\kappa}_{\lambda}  d_{\kappa} |^{2} \delta(\omega - E_{\lambda})
\label{eqn2}
\end{equation}
Here excitations are summed over excitons, $\lambda$, which are composed of linear combination of single-particle transitions,
$\kappa$, with weights, $\bar{A}^{\kappa}_{\lambda}$, and energy, $E_{\lambda}$. All these excitonic quantities can be computed by solving the BSE.
It is important to note that only ``vertical'' or ``direct'' ($q= k_{v}-k_{c} = 0$) excitons are relevant for the optical absorption. On the other hand, finite-momentum
excitons are particularly decisive for the emission profiles of indirect semiconductor systems such as some of the Ti$_2$CO$_2$ monolayer models studied
in this manuscript. Therefore, in these systems we solve the BSE also at the finite $q$ corresponding to their indirect band gaps (as reported in Tab.~\ref{table1}) in order to gain insight on their optical emission features.
\begin{table}[h!]
\caption{Energies of direct (D) and indirect (ID) excitons of the O-terminated Ti$_2$CO$_2$ monolayers with their respective binding energies in parentheses. All values are in eV.}\label{table2}
\begin{threeparttable}
\begin{ruledtabular}
\begin{tabular}{lcccc}
System & $\rm D_{1}$ & $\rm D_{2}$ & ${\rm ID}_{1}$ & ${\rm ID}_{2}$ \\
\hline
\vspace{-0.3cm}
\\
O-hTi& 1.39 (0.51) & 2.45 (0.76) & 0.80 (0.49) & --  \\
\hline
\vspace{-0.28cm}
\\
O-hC & 0.79 (0.98) & 1.46 (0.77) & 0.63 (0.56) & 0.67 (0.52)   \\
\hline
\vspace{-0.28cm}
\\
O-Ti & 1.20 (0.84) & -- & -- & --   \\
\hline
\vspace{-0.28cm}
\\
O-hTiC& 2.02 (0.73) & -- & 1.15 (0.59) & --
\end{tabular}
\end{ruledtabular}
\end{threeparttable}
\end{table}

The imaginary part of the dielectric function of the Ti$_2$CO$_2$ monolayer models are shown in Fig.~\ref{fig4} which are calculated using the QP energies and LDA wave functions. Spectra of the monolayers on top of the LDA eigenvalues are provided in the supplementary material of the manuscript for comparison. The solid and dashed blue lines in each figure corresponds to the spectrum with (GW+BSE) and without (GW+IPA) excitons, respectively. Note that plotted spectra correspond to ``vertical'' or ``direct'' ($q= k_{v}-k_{c} = 0$) excitons, we indicate the bound ``indirect'' excitons in the figures as dashed vertical lines where relevant. It is known that the exciton is a collective excitation, meaning that all electronic transitions in principle contribute to the excitonic peak, albeit with a certain weight. We provide the transitions that have the largest contribution to the specific excitonic peak in the BZ as subfigure for each monolayer. The binding energy of the excitons are reported in Tab.~\ref{table2} and calculated as the energy difference between the exciton energy and the QP single particle transition energy with greatest weight and same momentum.

\subsubsection{O-hTi monolayer}
O-hTi monolayer is an indirect band gap semiconductor with a band gap of 1.29 eV at the GW level (Tab.~\ref{table1}), where the VBM and CBM are at $\Gamma$ and $M$ points in the BZ, respectively (Fig.~\ref{fig2}(a)). Two prominent excitonic peaks D1 and D2 at 1.39 and 2.45 eV can be identified in the optical spectrum (Fig.~\ref{fig4}(a)) and among those peaks, D1 has a binding energy of 0.51 eV with corresponding transitons around $\Gamma$ point in the BZ as shown in the subfigure. The transitions which composed of the D2 exciton are, on the other hand, reside along the $\Gamma-M$ direction in the BZ as shown in the subfigure. Change in the wave function characteristics of the contributed orbitals  manifest itself in the binding energy of the D2 exciton which is calculated as 0.76 eV and rather larger than that of D1. In addition to these vertical excitons, we also indicate the energy, 0.80 eV, of the lowest energy indirect exciton (ID1) in the figure as a grey vertical dashed line. It is found that the ID1 exciton is the lowest energy exciton of the O-hTi monolayer with a binding energy of 0.49 eV for which the largest weight transitions reside between $\Gamma$ and $M$ points in the BZ. Upon comparison with the solar flux of terrestrial spectra, it is expected that O-hTi monolayer has strong absorption in the near infrared and visible (blue and green color) but indirect emission in the deep infrared region.

\subsubsection{O-hC monolayer}
Similar to O-hTi, O-hC monolayer is also an indirect band gap semiconductor with a band gap of 1.42 eV at the GW level (Tab.~\ref{table1}) where the VBM and CBM reside in between K-$\Gamma$ and $\Gamma$-$M$ directions, respectively (Fig.~\ref{fig2}(b)). We indicate the first direct excitonic peak in Fig.~\ref{fig4}(b) as D1 at 0.79 eV which originates from the transitions of the parallel bands between $\Gamma$-$M$ directions with a binding energy of 0.98 eV. After the D1 exciton in the absorption spectrum, there is a rather flat absorption region with several exctionic peaks originates from the same paralel bands between $\Gamma$-$M$ directions until D2 exciton at 1.46 eV. D2 peak has the largest oscillator strength in the low energy region which coincides with the near infrared region whose binding energy is 0.77 eV. Our finite-q BSE simulations showed that there are two indirect bound exciton, ID1 and ID2, at 0.63 eV and 0.67 with a binding energy of 0.56 and 0.52 eV, respectively. Upon comparison with the solar flux of terrestrial spectra, it is expected that
O-hC monolayer has strong absorption in the deep and near infrared but indirect emission in the deep infrared regime.

\subsubsection{O-Ti monolayer}
O-Ti monolayer is the only model which has a direct band gap (1.92 eV at the GW level) among the four Ti$_2$CO$_2$ Mxene monolayer models studied in this manuscript. The absorption spectrum shows one large peak, D1, which originates from direct transitions around $M$ point in the BZ at 1.20 eV and has a binding energy of 0.72 eV. Comparing with the solar flux of terrestrial spectra reveals that O-Ti monolayer has strong absorption and emission in the near infrared region. Being a direct band gap semiconductor with an optical gap in the infrared region signifies the potential usage of O-Ti monolayer in the infrared laser applications.

\subsubsection{O-hTiC monolayer}
O-hTiC monolayer is the other indirect Ti$_2$CO$_2$ Mxene monolayer model with a band gap of 1.74 eV at the GW level where the CBM and VBM are at $M$ and $\Gamma$ points in the BZ,
respectively. The absorption spectrum has one prominent excitonic peak, D1, at 2.02 eV with a binding energy of 0.73 eV. Our finite-q BSE simulations revealed the existence
of an indirect exciton, ID1, at 1.15 eV, which has a binding energy of 0.59 eV. Comparison with the solar flux of terrestrial spectra reveals that O-hTiC monolayer has
a strong absorption in the visible region (orange-yellow) and indirect emission in the infrared region.

\section{Conclusions}\label{SunCon}
In this manuscript, we present a state-of-the-art first principles analysis, based on DFT, GW and BSE formalisms, on the electronic and optical properties of four O-terminated Ti$_2$CO$_2$ MXene monolayers. We show that the electronic band gap of the monolayer models increase enormously upon inclusion of the GW correction compared to the LDA values. Using the GW-corrected QP energies and DFT wave functions, we then solve the BSE in order to investigate the direct and indirect excitons in these monolayers. We show that the absorption spectra of the monolayer models drastically redshifted upon inclusion of the electron-hole interaction due to the large binding energy of the excitons. We also observe that the binding energy of the indirect excitons are in general lower than the direct ones, however, they are still the lowest lying excitons in the absorption spectra of the indirect band gap monolayer models. We find that, despite some of them being strong absorbers in the visible region, they all likely are infrared emitters which opens the possibility of their usage in infrared laser and medical applications. Our findings in this manuscript emphasize the possible usage of surface termination to tune the optical and electronic properties of O-terminated monolayer models as well as the importance of inclusion of the many-body effects for the accurate prediction of the electronic and optical properties of 2D MXenes in general.
\begin{acknowledgements}
This material is based upon work supported by the Air Force Office of Scientific Research under award number FA9550-19-1-7048. Computing resources used in this work were provided by the  National Center for High Performance Computing of Turkey (UHeM) under grant number 1007502020. The numerical calculations reported in this paper were partially performed at TUBITAK ULAKBIM, High Performance and Grid Computing Center (TRUBA resources). CY acknowledges the computational support provided by the Computational Shared Facility at The University of Manchester. FP acknowledges the funding received from the European Union project MaX {\em Materials design at the eXascale} H2020-INFRAEDI-2018-2020/H2020-INFRAEDI-2018-1, Grant agreement n. 824143.
\end{acknowledgements}
\bibliography{references.bib}

\end{document}